\documentclass[prl,aps,11pt,preprintnumbers,longbibliography,superscriptaddress]{revtex4-1}

\usepackage{amssymb,bm}
\usepackage{graphicx}
\usepackage{color,xcolor}
\usepackage{amsmath} 
\usepackage{setspace}
\usepackage{enumerate}
\usepackage[bookmarks, colorlinks=true, breaklinks]{hyperref}
\hypersetup{linkcolor=blue,citecolor=blue,filecolor=black,urlcolor=blue}

\newcommand{\bra}{\left< }
\newcommand{\ket}{\right>}
\newcommand{\ovl}[1]{\overline{#1}}

\DeclareMathOperator{\DM}{\mathcal{D}}

\emergencystretch=\maxdimen
\begin{document}

\title{Random electric field instabilities of relaxor ferroelectrics}

\author{Jos\'{e} R. Arce-Gamboa} 
\affiliation{Centro de Investigaci\'{o}n en Ciencia e Ingenier\'{i}a de Materiales and Escuela de F\'{i}sica, Universidad de Costa Rica, San Jos\'{e}, Costa Rica 11501,}

\author{Gian G. Guzm\'{a}n-Verri\footnote{gian.guzman@ucr.ac.cr}  } 
\affiliation{Centro de Investigaci\'{o}n en Ciencia e Ingenier\'{i}a de Materiales and Escuela de F\'{i}sica, Universidad de Costa Rica, San Jos\'{e}, Costa Rica 11501,}
\affiliation{Materials Science Division, Argonne National Laboratory, Argonne, Illinois, USA 60439}

\date{\today}

\begin{abstract}

Relaxor ferroelectrics are complex oxide materials which are rather unique to study the effects of compositional disorder on phase transitions. Here, we study the effects of quenched cubic random electric fields on the lattice instabilities that lead to a ferroelectric transition and show that, within a microscopic model and a statistical mechanical solution, even weak compositional disorder can prohibit the development of long-range order and that a random field state with anisotropic and power-law correlations of polarization emerges from the combined effect of their characteristic dipole forces and their inherent charge disorder. We compare and reproduce several key experimental observations in the well- studied relaxor PbMg$_{1/3}$Nb$_{2/3}$O$_3$-PbTiO$_3$.

\end{abstract}

\maketitle

\section{Introduction}

Relaxors exhibit  a myriad of complex phenomena that 
are both scientifically interesting and technologically important such as
diffuse phase transitions where large and frequency dependent dielectric permittivities extend over  
 hundreds of Kelvin degrees~\cite{Bovtun2004a} without any signature of macroscopic symmetry breaking as well as unltrahigh electromechanical
responses~\cite{Park1997a, Guo2003a, Manley2016a, Li2016a}. These properties make relaxors  attractive material 
candidates for energy storage and harvesting applications 
as well as future cooling technologies for integrated microelectronics~\cite{Uchino2015a, Moya2014a, Scott2007a}.  

Though relaxors were first synthesised in the 1950s~\cite{Rustum1954a, Smilenskii1950a} and 
they have been the subject of many theoretical~\cite{Pirc1999a, Tinte2006a, Ganesh2010a,  Takenaka2013a, Al-Barakaty2015a} and experimental 
studies~\cite{Al-Zein2010a, Bosak2012a, Phelan2014a,  Chernyshov2015a, Phelan2015a, Hehlen2016a, Manley2014a, Kleemann2016a} 
there is still no consensus on a satisfactory theory of relaxor ferroelectricity~\cite{Cowley2011a}.  
One of the major difficulties in describing relaxors is that they exhibit many
characteristic temperatures. From high to low, these are (i)
the Burns temperature $T_B$ below which its dielectric response
deviates from Curie-Weiss law behavior with (ii) a corresponding Curie-Weiss temperature $T_{CW}$;
(iii) a frequency  dependent temperature $T_{max} $ where the susceptibility is maximum but no ferroelectric (FE) transition occurs; and (iv) an induced FE transition temperature $T_c $ if sufficiently large electric fields are applied. Crucially, X-ray and neutron scattering studies have found  anisotropic quasi-elastic diffuse scattering very near $T_{CW}$~\cite{Xu2004a, Gvasaliya2005a, Stock2007a, Rotaru2008a, Gehring2009a}.

It has been recognized that a central question in the discussion of relaxors is the effect of random electric fields 
on the FE transition of cubic systems such as the typical perovskite relaxor PbMg$_{1/3}$Nb$_{2/3}$O$_3$~(PMN)~\cite{Westphal1992a, Cowley2011a, Phelan2014a}.
The random electric fields  originate from charge disorder:  cations with different charge valencies are randomly located on the octahedrally coordinated site such as 
Mg${^2+}$ and Nb$^{5+}$  in PMN~\cite{Westphal1992a}. These ions do not order with temperature, making the compositional disorder quenched.  
Unlike the widely studied random fields (RFs) in magnets  which linearly couple to an order parameter of the Ising or Heisenberg type~\cite{Imry1975a}, 
the quenched electric RFs of relaxors couple to a cubic order parameter~\cite{Cowley2011a}. 
It is believed that $T^{\ast}$ is the onset temperature of a RF state in which relaxors exist~\cite{Cowley2011a}.

In addition to the symmetry of the order parameter, 
we make the observation that the characteristic dipolar interaction of FEs is equally important. 
It is well-known that the structural instability that leads to the breaking of lattice inversion symmetry 
and a spontaneous polarization, is the result of dipolar forces between electric dipole moments 
induced by the displacements of the ions associated with a zone-center transverse optic (TO) mode~\cite{Lines1997a}. 
Such dipolar forces are highly anisotropic and long-ranged, which are very much in contrast with 
the isotropic and short-ranged exchange couplings between the spin degrees of freedom of magnets. 
According to the theory of phase transitions~\cite{Goldenfeld1992a}, FEs and magnets are therefore in 
different universality classes, rendering the  standard models that describe the effects of 
RFs on magnetic transitions~\cite{Imry1975a} inadequate for relaxors~\cite{Cowley2011a}.

In a previous paper,~\cite{GuzmanVerri2013a} we studied the effects on quenched electric RFs in a standard, uniaxial displacive model of the FE transition.  Within a statistical mechanical variational solution, we showed that intrinsic polarization fluctuations associated with the dipolar force and RF disorder, result in diffuse phase transitions - a hallmark of relaxor behavior. Typical relaxors such as PMN are cubic, however, and there is no a-priori reason to believe that the results for uniaxial systems will hold in environments with higher symmetries.  The purpose of this work is then to study the random electric field problem posed by cubic relaxors within a minimal microscopic model. We extend the uniaxial model Hamiltonian of Ref. [\onlinecite{GuzmanVerri2013a}] to cubic symmetries 
by including the usual displacement soft-mode coordinates along each cubic axis, cubic anisotropy, dipole tensor, and cubic RFs. We also extend to cubic symmetries our previously developed variational solution for uniaxial systems.
We will show that as a result of the combined effect of dipolar forces and quenched RFs
a state with no-long range FE order and anisotropic, long-ranged fluctuations of  polarization
emerges for any amount of compositional disorder. We identify this disordered state as the RF state of relaxors.  
We will also show that long-ranged FE order can be induced by application of strong enough electric fields and that
such transition ends at a critical point, as it is observed in experiments~\cite{Kutnjak2006a}.

\section{Results}

We consider a cubic lattice  
and choose normal mode coordinates that describe local displacements 
$( Q_{ix}, Q_{iy}, Q_{iz})$   in the unit cell $i$  that are 
associated with the soft TO mode, the condensation of which
leads to the FE transition~\cite{Pytte1972a}. We consider the model Hamiltonian,

\begin{align}
\label{eq:Hamiltonian}
H &=    \frac{1}{2} \sum_{i \lambda }  \Pi_{i \lambda }^2 + \frac{\kappa}{2} \sum_{ i \lambda  }  Q_{ i \lambda }^2
+ \frac{\gamma_1}{4} \sum_{ i \lambda  }  Q_{ i \lambda }^4   + \frac{ \gamma_2 }{ 4 } \sum_{i, \lambda \neq \lambda^\prime }  Q_{ i \lambda }^2  Q_{ i \lambda^\prime }^2 \nonumber \\
& \hspace{6cm} 
-\frac{1}{2} \sum_{ij \lambda \lambda^\prime} v_{ij}^{\lambda \lambda^\prime}  Q_{ i \lambda }   Q_{ j \lambda^\prime }  - \sum_{i \lambda } E_{ \lambda }^0 Q_{ i \lambda } 
- \sum_{i \lambda } h_{ i \lambda } Q_{ i \lambda },
\end{align}
with  $\lambda,\lambda^\prime = x,y,z$. $ \Pi_{i \lambda } $ is  
the conjugate momentum of $ Q_{ i \lambda } $; $  E_{\lambda }^0  $ is an applied electric field; and 
$ v_{ij}^{\lambda \lambda^\prime} $ is the dipolar interaction tensor with 
Fourier transform
$
v_{\bm q}^{ \lambda \lambda^\prime } 
=  \left[ \frac{1}{3}C^2 - B^2 |{\bm q}|^2  \right] \delta_{ \lambda \lambda^\prime } -C^2 \frac{ q_{ \lambda } q_{ \lambda^\prime } }{ |{\bm q}|^2 },  
$
where $ |{\bm q}| = \sqrt{q_x^2 + q_y^2  + q_z^2}$ is the magnitude of the wavevector ${\bm q}$; and  $B$ and $C$ are constants that depend on the lattice structure~\cite{Aharony1973a}. 
Hereafter, we denote $v_0 = C^2/3 $ as the component of $v_{\bm q}^{ \lambda \lambda } $ when $ {\bm q} \to 0$ in the direction transverse to $\lambda$ (the value of $v_{\bm q}$ depends on the direction in which ${\bm q}$ approaches zero). $\kappa$ is the lattice stiffness and $ \gamma_{1,2}$ are anharmonic coefficients. 

For the quenched random fields $h_{i \lambda}$, we choose a Gaussian probability distribution of independent random variables 
 with zero mean and variance $\Delta^2$.
In the absence of compositional disorder, this is a standard minimal model for ferroelectricity in cubic perovskites~\cite{Pytte1972a}.

To study the statistical mechanics of the Hamiltonian (\ref{eq:Hamiltonian}), it is necessary to 
consider thermal and quantum fluctuations at least at the level of the Onsager approximation  
and random field fluctuations at least at the level of a replica theory~\cite{GuzmanVerri2015a}.   
To do so,  we generalize 
a variational method previously developed by one of us~\cite{GuzmanVerri2013a} 
to cubic symmetries. 
Such method allow us to calculate the temperature and disorder dependence of  
relevant quantities such as the phonon frequencies, the  polarization order parameter and the correlation functions in a self-consistent fashion. 
The details are presented in Methods section. 

Our model parameters are $ \kappa, \gamma_1, \gamma_2, v_0, B $ and $\Delta$.  
Throughout this work, we have  fixed the values of $ \kappa, \gamma_1, \gamma_2, v_0,$ and $B$ 
 to those of typical values of oxide perovskites~\cite{Pytte1969a}  
and  to fit the transition temperature of the conventional FE PbTiO$_3$ (PTO, $T_c^0 \simeq 760\,$K )~\cite{Landolt-Bornstein} 
assuming $\Delta=0$.
The resulting values are given in Table \ref{t:parameters}.
Our choice gives  a Curie-Weiss constant of  $C_{CW}^0 \simeq 2.4 \times 10^{5}\,$K and a zone-center
TO phonon energy of $\Omega_0^\perp \simeq 5.2\,$meV at zero temperature,
which are typical of conventional FEs.  
Depending on the choice of the anharmonic coefficients, the low temperature FE phase predicted by the Hamiltonian (\ref{eq:Hamiltonian}) in the absence of compositional 
disorder has tetragonal ($ \gamma_1 < \gamma_2 $) or rhombohedral ($ \gamma_1 > \gamma_2 $) symmetry~\cite{Pytte1972a}.
In this work, we have chosen  $\gamma_1 > \gamma_2$, as we will study the field-induced FE transition of relaxors, which is 
typically a cubic-to-rhombohedral structural phase change~\cite{Phelan2015a}.

\begin{table}[h]
\caption{Model parameters used in this work.}
\centering
\begin{tabular}{cccccc} \hline \hline
    $\omega_0^2 \equiv v_0 - \kappa $(meV$^2$)    & $ B^2 $(meV$^2$ \AA$^2$)  &  $ \gamma_1 $(meV$^3$)  & $  \gamma_2 $(meV$^3$)  & $v_0\,(=C^2/3) $(meV$^2$)  \\ \hline
        $  21  $                            &   $ 3500 $                          &     $ 272 $                                   &   $ 200 $                                  & $ 5071$                        \\  \hline \hline
\end{tabular}
\label{t:parameters}
\end{table}

\begin{figure}[htp]
\centering
\includegraphics[scale=0.4]{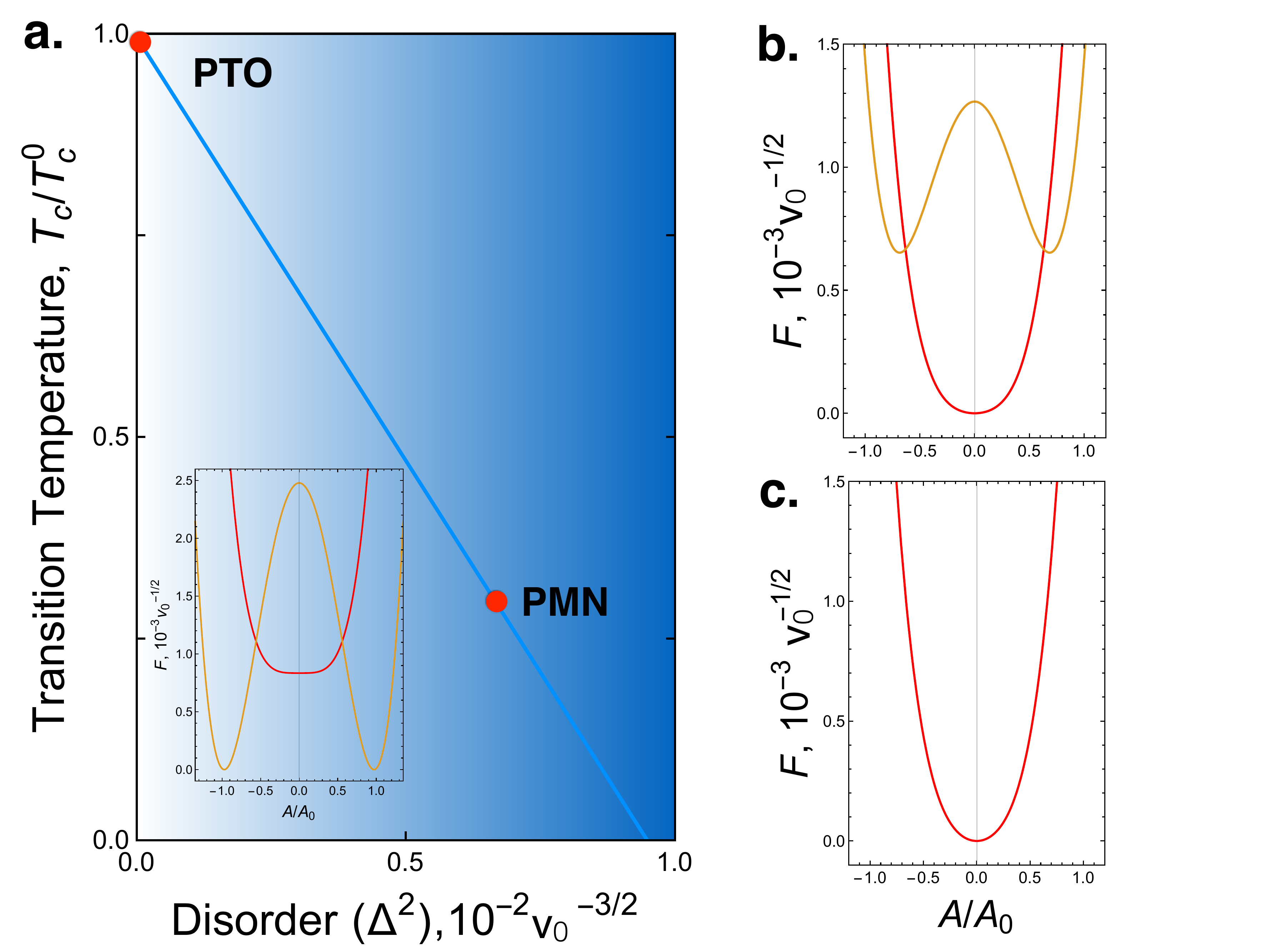}
\caption{Temperature-disorder phase diagram and free energies.  (\textbf{a}) The RF state is stable above the FE transition line 
and becomes metastable below it down to zero temperature. The inset shows the free energies of the RF (red) and FE (yellow) states at zero temperature for
 $\Delta^2/v_0^{3/2}= 0.5 \times10^{-2}$. (\textbf{b})-(\textbf{c}) Zero temperature free energies for moderate ($\Delta^2/v_0^{3/2}=2.0 \times 10^{-2})$ and strong ($3.0 \times 10^{-2}$) compositional disorder, respectively,
 showing the RF state as a global minimum.  $A_0$ is the order parameter for the pure case at zero temperature. All energies are plotted with respect to their corresponding minimum.}
\label{fig:phasediagram}
\end{figure}

We first present our results in the absence of applied electric fields. 
Figure \ref{fig:phasediagram} shows the calculated temperature-disorder phase diagram and the zero temperature free energies where we have identified three regions according to
the RF strength. For weak RFs ($0 \leq \Delta^2/v_0^{3/2}  \lesssim 0.9$),  long-range FE order sets in at a transition temperature 
$T_c < T_c^0$ and it is  accompanied by a metastable random field disordered state down to $T=0$, as it is shown in  
Fig.\,\ref{fig:phasediagram}\,(a).
For moderate compositional disorder ($0.9 \lesssim \Delta^2/v_0^{3/2}  \lesssim 2.2$), there is no transition as the the 
RF state becomes stable at all temperatures and the long-ranged polar state is
now metastable, see Fig.\,\ref{fig:phasediagram}\,(b).  For strong compositional disorder ($ \Delta^2/v_0^{3/2}  \gtrsim 2.2 $), only the RF state exists, as it is shown in 
Fig.\,\ref{fig:phasediagram}\,(c).

By comparing our phase diagram with that of the relaxor PbMg$_{1/3}$Nb$_{2/3}$O$_3$-PbTiO$_3$ (PMN-PT)~\cite{Phelan2015a} and 
assuming that the conventional FE PTO is near about $\Delta=0$ with $T_c/T_c^0 \simeq 1$, 
then PMN is in the weak-disorder region with $T_c/T_c^0 \simeq 0.3$ and $\Delta^2/v_0^{3/2} \simeq 0.7  \times 10^{-2}$, as it is shown in Fig.\,\ref{fig:phasediagram}\,(a).
This means that while the ground state of PMN-PT relaxors is FE, 
those in the Ti-poor side of the morphotropic phase boundary
are stuck in a metastable disordered random field state below the phase transition line.
We will see below that this is also supported by the predicted correlation lengths and static susceptibilities of our model.

\begin{figure}[htp]
\centering
\includegraphics[scale=0.7]{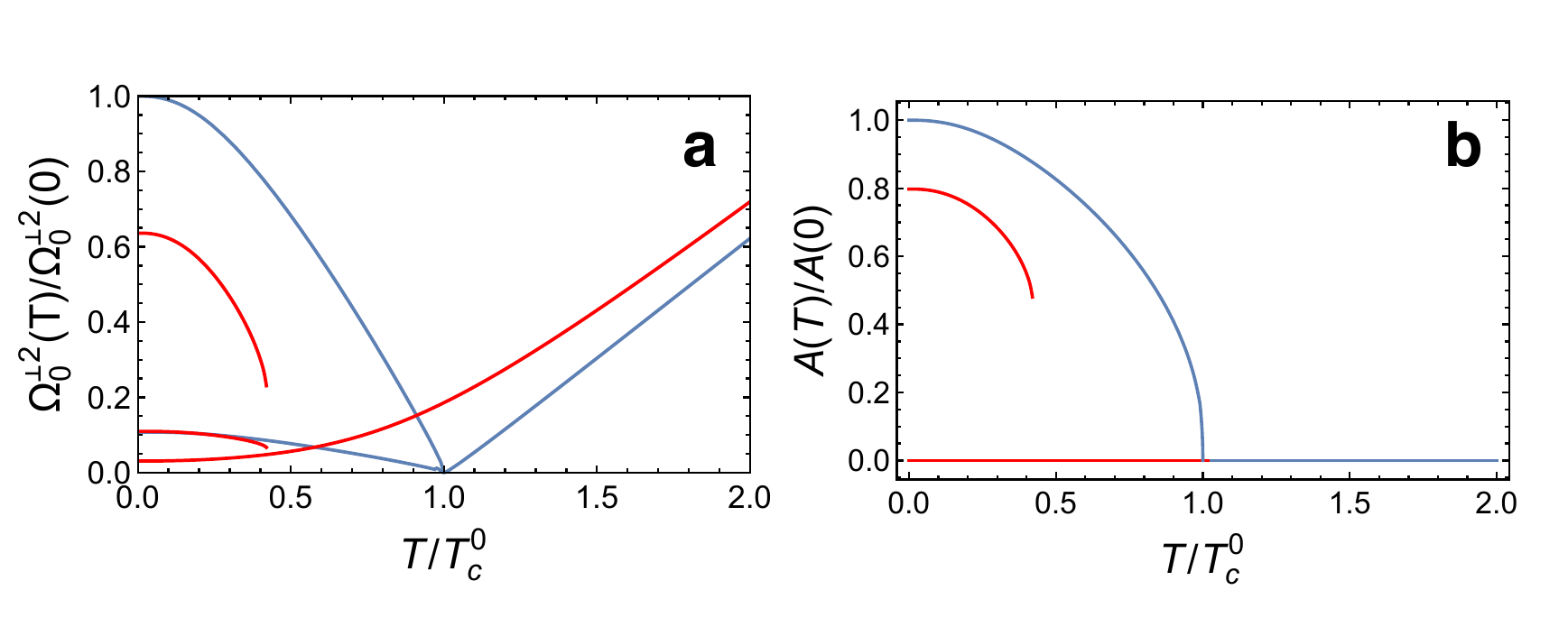}
\caption{Phonon frequencies and order parameter. Temperature and disorder dependence of (\textbf{a}) the squared of the phonon frequencies and 
 (\textbf{a}) spontaneous polarization along $(111)$. Here, $ \Delta^2 / v_0^{3/2} = 0\, (\mbox{blue}), 1.5 \times 10^{-2}\, (\mbox{red})$ and 
  $\Omega_0^\perp (0)$ and $A(0)$  are the TO frequency and order parameter of the pure FE at zero temperature. }
\label{fig:polarization}
\end{figure}

The temperature and disorder dependence of the zone-center TO phonon frequency, $\Omega_0^{\perp}$, and the order parameter, $A$,  
associated with the RF and FE states are shown in Figure \ref{fig:polarization} (a) and (b), respectively. 
While the TO mode softens and condenses at $T_c^0$ for the pure case, as expected,
that of the RF state remains finite all the way down to zero temperature for any amount of compositional disorder.
The temperature dependence of the metastable states is shown for the sake of completeness. 
When contrasted to experiments~\cite{Hehlen2016a}, the observed softening of the phonon frequency of the RF state  
above about $T^*$ is in qualitative agreement with our model and we will show that it supports the conclusion that such softening is responsible for the large increase observed in the dielectric constant. Below $T^*$, however, the observed frequencies exhibit a more complex behavior not captured by our model.  
We believe that some of the discrepancies are due to local spontaneous polarizations in the disordered state, which we do not allow in our model.

We now discuss the correlation functions in the fluctuations of polarization of the RF state. 
As it is usually done for conventional cubic FEs~\cite{Strukov98a}, we consider mean squared fluctuations 
on the polarization components $Q_{ {\bm q} }^\perp$ and $Q_{ {\bm q} }^\parallel$ that are transverse and longitudinal to a wave-vector ${\bm q}$, respectively. 
For the transverse components we obtain isotropic fluctuations with the following form,
\begin{align}
\label{eq:SqTransverse}
\ovl{ \bra {Q_{ {\bm q} }^\perp }^2  \ket } =  \frac{ 1 }{ 2  \Omega_{ {\bm q} }^\perp   } \coth\left( \frac{ \beta  \Omega_{ {\bm q}}^\perp }{2} \right) + \frac{\Delta^2}{ { \Omega_{ {\bm q} }^\perp}^4},
\end{align}
where $ {\Omega_{ {\bm q} }^\perp}^2  = B^2 \left( \xi^{-2} + \left| {\bm q} \right|^2 \right)$ is the doubly degenerate TO mode (see Methods section) and  where we have identified $\xi =  B / \Omega_{0}^\perp$ as the correlation length. 
$\ovl{\bra \dots \ket}$ denotes thermal and compositional averages taken in that order.
In the absence of disorder and in the classical limit~($ \beta  \Omega_{ {\bm q}\lambda} \ll 1 $), 
Eq.~(\ref{eq:SqTransverse}) reproduces to the fluctuations of pure FEs~\cite{Strukov98a}.
In the classical limit, Eq.~(\ref{eq:SqTransverse}) becomes a Lorentzian plus a Lorentzian squared. 
While this is analogous to the well-known result of the random field Ising model \cite{Young1991a}, we will show below that 
the correlation functions behave very differently in real space due to the anisotropy and long-range nature of the dipole force.
The wave-vector distribution predicted by  Eq.~(\ref{eq:SqTransverse}) has been recently observed in diffuse scattering experiments \cite{Chernyshov2015a}
and the quantum fluctuations have been found important to correctly describe the observed static structure factor at low temperatures~\cite{GuzmanVerri2015a}.
For the fluctuations in $Q_{ {\bm q} }^\parallel$, we find that they have a similar form to that of Eq.~(\ref{eq:SqTransverse}) except that 
 the TO frequency is replaced by that of the the longitudinal mode ${ \Omega_{ {\bm q} }^\parallel}^2 \propto  \xi^{-2} + \left| {\bm q} \right|^2 + C^2$. The constant $C$
is related to the depolarizing field~\cite{Strukov98a}, which makes these fluctuations 
significantly smaller than those in $Q_{ {\bm q} }^\perp$.

\begin{figure}[htp]
\centering
\includegraphics[scale=0.5]{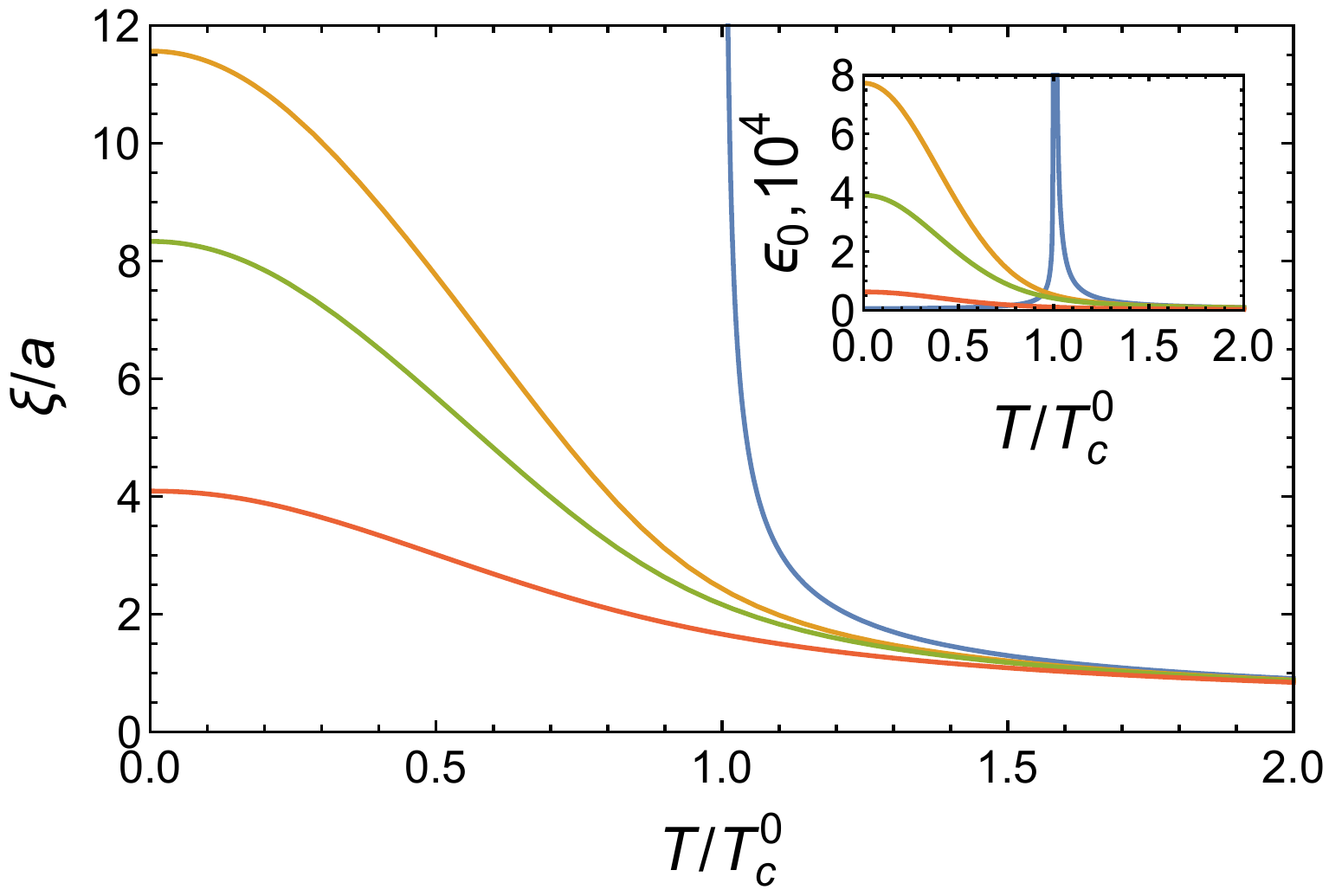}
\caption{Correlation length of fluctuations of polarization.  Temperature dependence of the correlation length and (inset) static dielectric constant for several 
random field strengths. Here, $ \Delta^2 / v_0^{3/2} = 0.0 \, (\mbox{blue}), 0.5 \times 10^{-2} \, (\mbox{orange}),  0.7 \times 10^{-2} \, (\mbox{green}),  1.5 \times 10^{-2} \, (\mbox{red})$. }
\label{fig:correlationlength}
\end{figure}

Figure \ref{fig:correlationlength} shows the calculated temperature dependence of the correlation length $\xi$ of the RF state for several disorder strengths.
In the absence of disorder, $\xi$ diverges as  expected near the FE transition. In the presence of disorder, the correlation length of the random field state 
remains finite at all temperatures. At $T=0$  it scales with disorder as $\xi \propto 1/\Delta^2$, which we identify as the minimum length scale on which domains must appear spontaneously. 
By a standard procedure~\cite{Strukov98a}, it can be shown  that the static dielectric susceptibility is given by $\chi = \frac{3}{4\pi}\frac{v_0}{B^2}\xi^2$. 
The temperature and disorder dependence of the resulting static dielectric constant $\epsilon=1+4\pi \chi $, are shown in the inset of Fig.\,\ref{fig:correlationlength}.
We find that our model is in fair qualitative and quantitative agreement with the measured correlation length~\cite{Gvasaliya2005a}  
and static dielectric constant~\cite{Levstik1998a} in PMN when  $ 0.5 \lesssim \Delta^2/v_0^{3/2} \lesssim 0.7 $. This is  consistent with our identification
of the Ti-poor region of PMN-PT in our phase diagram (see Fig.\,\ref{fig:phasediagram})  and where the RF state is metastable.
\begin{figure}[htp]
\centering
\includegraphics[scale=0.7]{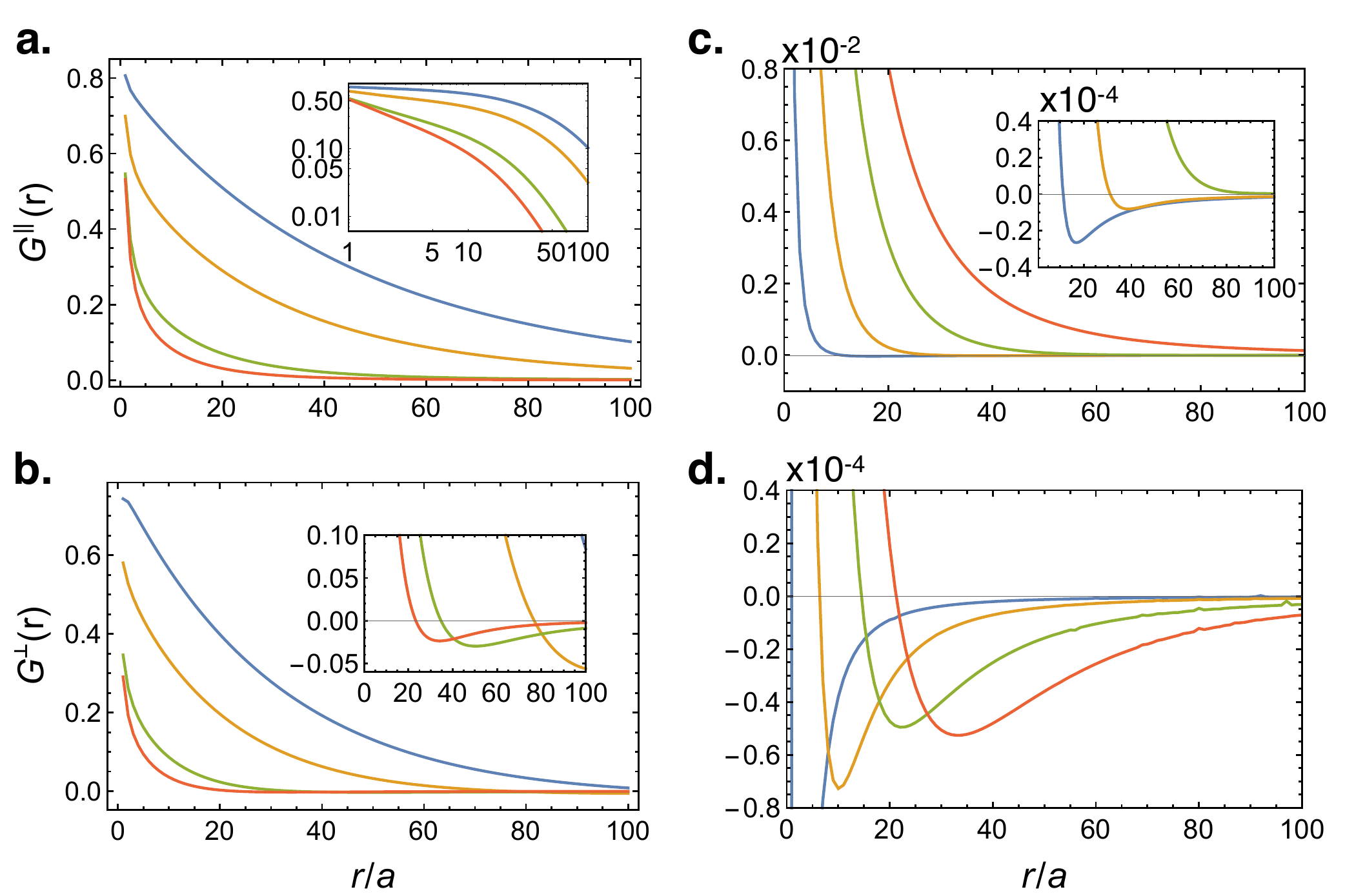}
\caption{Correlation functions of polarization.  Spatial dependence of the longitudinal and transverse components 
of the correlation functions (\textbf{a-b})  with  and without  ($\textbf{c-d}$) disorder ($\Delta^2/v_0^{3/2}=0.7\times 10^{-2}$) at several
temperatures. Here, $T/T_c^0 = 0.0  \, (\mbox{blue}), 0.5  \, (\mbox{yellow}), 0.9   \, (\mbox{green}), 1.1   \, (\mbox{red}) $. 
The inset in (a) is a log-log plot of the longitudinal correlations showing the crossover from exponential to power-law behavior;
and the insets in (c) and (d) show the change in sign in the correlations with increasing distance. }
\label{fig:CFs}
\end{figure}
We now discuss the spatial dependence of the correlation functions of polarization $ G_{\lambda \lambda^\prime}({\bm r})$.
Figure \ref{fig:CFs}  shows $ G_{\lambda \lambda}({\bm r})$ calculated from Fourier transform of Eq. (\ref{eq:SqTransverse}).
In the presence of compositional disorder, the correlations of the random field state are positive along the 
longitudinal direction ($\bm{r} \parallel \bm{\lambda}$) and they change sign 
in the transverse direction  ($\bm{r} \perp \bm{\lambda} $), as it is shown in Figs. \ref{fig:CFs}\,(a) and \ref{fig:CFs}\,(b), respectively. 
For short distances compared to the correlation length $\xi$,  they fall-off exponentially 
and  then cross over to a power law behavior ($\propto r^{-3}$) for $ r \gg \xi$,  see inset in Fig.\,\ref{fig:CFs}\,(a).
We verify the large distance behavior by calculating analytic expressions of the correlation functions of the random field state  in the classical limit,
\begin{align*}
   G_{\lambda \lambda}(\bm{r})  &= \begin{cases} 
~~\frac{4\pi^2 \xi}{v_{BZ}B^2} \left( \frac{k_B T}{\xi^2} + \frac{\Delta^2}{B^2}  \right)  \left( \frac{r}{\xi} \right)^{-3} + \mathcal{O}\left( e^{- r / \xi}  \right) , & \bm{r} \parallel \bm{\lambda}, \\
-\frac{2\pi^2 \xi}{v_{BZ}B^2} \left( \frac{k_B T}{\xi^2}  + \frac{\Delta^2}{B^2}  \right)  \left( \frac{r}{\xi} \right)^{-3} + \mathcal{O}\left( e^{- r / \xi}  \right), & \bm{r} \perp \bm{\lambda},
\end{cases}, \\
G_{\lambda \lambda^{\prime}}( \bm{r} ) &= 0, ~~~~\lambda \neq \lambda^{\prime},
 \end{align*}
where $\lambda=x,y,z,$ and $v_{BZ}$ is the volume of the Brillouin zone.  Note that the corrections to these power laws are exponentially small. 
The cross-component correlations ($\lambda \neq \lambda^\prime$) of the random field state are identically zero everywhere, 
as expected from cubic symmetry.
 
Note they also increase with decreasing temperature but do not reach long-range order as their correlation length $\xi$ remains finite 
for all temperatures. 
This is in stark contrast with  the correlations of the pure compound  where, while anisotropic, they are strongest near the FE transition and then 
weaken away from it, as it is shown in Figs. \ref{fig:CFs}\,(c) and \ref{fig:CFs}\,(d).
Previous theoretical work have also found anisotropic correlations~\cite{Takenaka2013a, Al-Barakaty2015a}.

\begin{figure}[htp]
\centering
\includegraphics[scale=0.6]{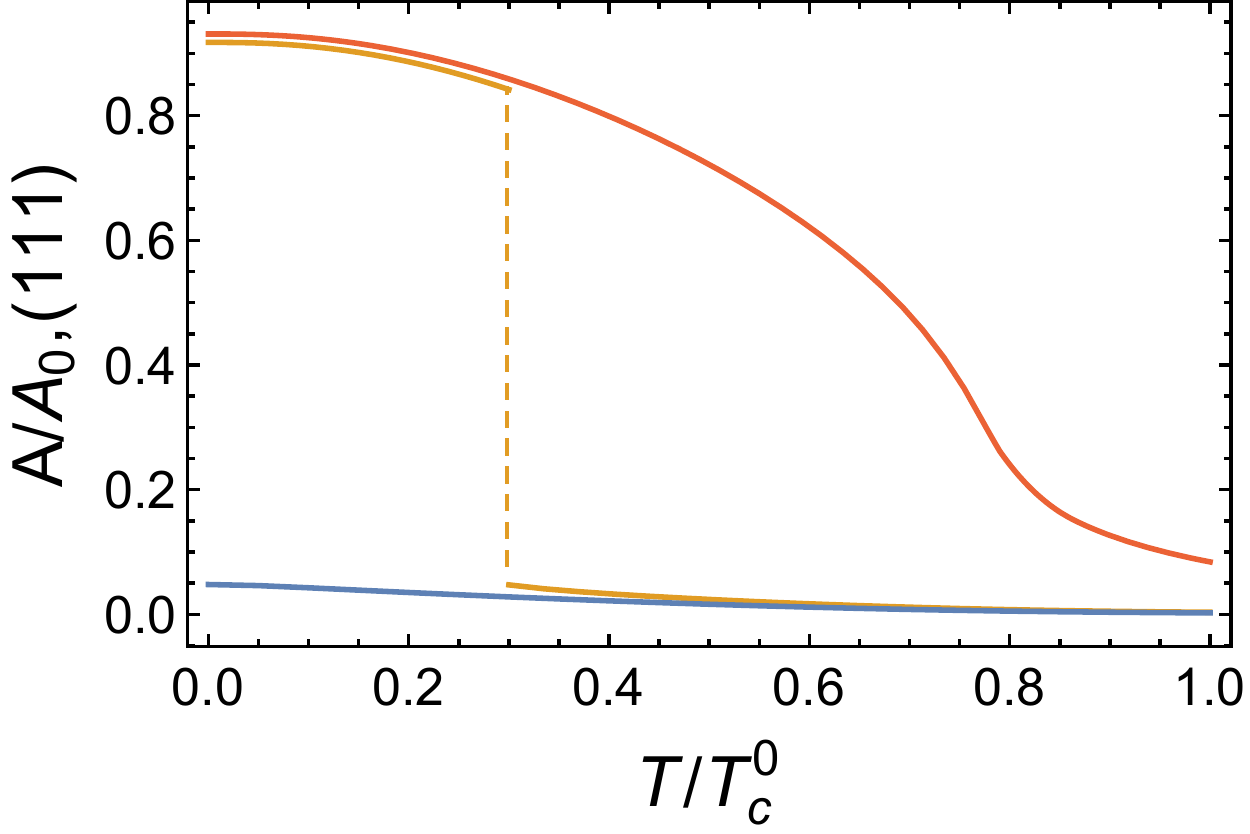}
\caption{Field-induced FE transition. Temperature dependence of the order parameter for several applied electric fields.  
 A relaxor-to-FE transition is induced for moderate electric fields 
 ($ E_0 / \Delta = 0.5 \times 10^{-3} $, orange line).
 Upon increasing the field strength ($ E_0 / \Delta =  1.2 \times 10^{-2} $, red line)  the system approaches a critical point, as observed in experiments~\cite{Kutnjak2006a}.
Here $ \Delta^2 / v_0^{3/2} = 0.7 \times 10^{-2} $ and $A_0$ is the order parameter at zero temperature for the pure compound.}
\label{fig:PMNPolarization}
\end{figure}

We now describe our results in the presence of an applied electric field.  Figure \ref{fig:PMNPolarization}, shows the temperature dependence of the order parameter for weak disorder and several field strengths. 
For weak applied electric fields $ \left( 0 < E_0 / \Delta \lesssim 0.4 \times 10^{-3} \right) $, the polarization of the RF state grows with decreasing temperature without inducing a FE transition. 
For moderate field strengths $ \left( 0.4 \times 10^{-3} \lesssim E_0/ \Delta \lesssim  1.0 \times 10^{-2}  \right) $, a clear first-order transition occurs as shown by the discontinuity in the order parameter. 
This discontinuity becomes weaker with increasing applied field until it reaches a critical point 
where the transition is  of second order. For strong applied fields $\left( E_0 / \Delta \gtrsim  1.0 \times 10^{-2}  \right)$, the transition is smeared. 
This behaviour is in agreement with experiments in PMN-PT~\cite{Kutnjak2006a}.

\section{Discussion}
We now compare our results to previous theoretical works. When contrasted to uniaxial systems~\cite{GuzmanVerri2013a, GuzmanVerri2015a}, we find that they share some similarities at the qualitative level such as the emergence of a RF state with an energy gap from the FE ground state and field-induced transitions that end at a critical point. The most significant difference appears in the correlation functions of polarization at short distances, where there is no partial screening of dipoles in the uniaxial case. Instead, the power-law  tails join smoothly to a short-range part where they saturate to near the on-site correlations. Our results disagree with the work of Sherrington \cite{Sherrington2014a}, where heuristic arguments are given to conclude that relaxor behavior in heterovalent compounds is mainly due to bond disorder and that RFs only play a secondary role. On the other hand, our results support the view of Takenaka et al.~\cite{Takenaka2013a} that there is no non-polar matrix in relaxors (our correlations decay as power-laws for $r \gg \xi$); and that of Al-Barakaty et al.~\cite{Al-Barakaty2015a} that quenched RF disorder is essential for relaxor behavior.  We emphasize, though, that, according to our results, the intrinsic fluctuations associated with the concomitant dipole forces are essential as well.

To  summarize, we have studied the effects of cubic random electric fields on the lattice instabilities that lead to a 
 FE transition. We have shown that a RF state emerges from the
combined effect of dipolar forces and compositional disorder. Such state has no-long range FE order and anisotropic, long-ranged correlations 
of polarization that grow with decreasing temperature.  
When comparing to the experimental phase diagram of typical relaxors such as PMN-PT, we conclude that 
while the ground state is FE, the RF state in the poor Ti side of the
phase diagram is metastable below the phase transition line down to zero Kelvin. Upon application of strong enough electric fields, first-order transitions can be induced and end at a critical point. 
While we have focused our attention on PMN-PT, our model and results are generic and should be applicable 
to other relaxors such as PbZn$_{1/3}$Nb$_{2/3}$O$_3$-PbTiO$_3$. 

\section{Methods}
We now describe our variational solution of our model Hamiltonian of Eq. (\ref{eq:Hamiltonian}).
We consider the trial probability distribution, 
\begin{align}
\label{eq:rhotrial}
\rho^{tr}  = \frac{ 1 }{ Z^{tr} } e^{- \beta H^{tr} },  
\end{align}
where $H^{tr}$ is the Hamiltonian of a displaced, cubic harmonic oscillator in a random field $h_{i \lambda}$,
\begin{align*}
H^{tr} &=\frac{ 1 }{ 2 } \sum_{i \lambda }  \Pi_{i \lambda }^2 + \frac{ 1 }{ 2 } \sum_{ ij,\lambda \lambda^\prime } \left( Q_{i \lambda } - A_{i \lambda} \right)  ( \DM_{i-j} )^{ \lambda \lambda^\prime } \left( Q_{j \lambda^\prime} - A_{j \lambda^\prime} \right) 
- \sum_{ i \lambda} h_{i \lambda} Q_{i \lambda} , 
\end{align*}
and $Z^{tr}$ its normalization,
\begin{align*}
Z^{tr} = \mbox{Tr} e^{- \beta H^{tr} } 
& = \left( \prod_{ {\bm q}, \alpha }   \left[ 2 \sinh \left( \frac{  \beta \Omega_{ {\bm q} \alpha } }{2} \right) \right]^{-1} \right) 
 \times
 e^{\frac{\beta}{2} \sum_{ ij,\lambda \lambda^\prime }  h_{i \lambda }   ( \DM_{i-j}^{-1} )^{\lambda \lambda^\prime 	}  h_{j \lambda^\prime }  + \beta \sum_{i \lambda }  h_{i \lambda} A_{ i \lambda } },
\end{align*}
where  $  \Omega_{ {\bm q} \alpha }~(\alpha=1,2,3) $ are the soft mode frequencies at wave vector ${\bm q}$ and are given  by the squared of the the eigenvalues 
of the Fourier transform of the dynamical matrix  
$ (  \DM_{\bm q} )^{ \lambda \lambda^\prime }=\sum_{{\bm R}_{ij}}  (  \DM_{i-j} )^{ \lambda \lambda^\prime } e^{i {\bm q} \cdot {\bm R}_{ij}} $. $A_{i \lambda }$ is the $\lambda$-component of the  order parameter  at site $i$ and it corresponds to the mean displacement averaged over thermal and compositional disorder,
\begin{equation*}
A_{i \lambda } = \overline{ \bra Q_{i \lambda } \ket }= \int_{-\infty}^\infty \left( \prod_{i=1}^N \prod_{ \lambda =x,y,z } dh_{i\lambda} \right) P( \left\{ h_{i \lambda} \right\} ) 
\mbox{Tr} \left\{ \, \rho^{tr} \, Q_{i \lambda } \right\}.
\end{equation*}

We now compute the free energy $ \ovl{F} = \ovl{\bra H \ket} + T  \ovl{  \bra  k_B \ln \rho^{tr} \ket} $ using our probability distribution (\ref{eq:rhotrial}) together with the above equations.
The result is the following,
\begin{align}
\label{eq:free1}
\ovl{F} &= \frac{1}{4} \sum_{{\bm q} \lambda} \Omega_{{\bm q} \lambda} \coth\left( \frac{ \beta  \Omega_{ {\bm q} \lambda}}{2} \right)  +
\frac{\kappa}{2} \sum_{ i \lambda  }  \left[ A_{i\lambda}^2 +  \psi_{ii}^{\lambda \lambda}  \right] 
+ 
\frac{\gamma_1}{4} \sum_{i \lambda } \left[ 3 \left( \psi_{ii}^{ \lambda \lambda } \right)^2 
+
6 \psi_{ii}^{ \lambda \lambda }  A_{i \lambda}^2
+
A_{ i \lambda }^4   \right] \nonumber \\
&+ 
\frac{\gamma_2}{4}\sum_{i, \lambda  \neq \lambda^\prime }
\left[
2 \left( \psi_{ii}^{ \lambda \lambda^\prime } \right)^2 
+ \psi_{ii}^{ \lambda \lambda } \psi_{ii}^{ \lambda^\prime \lambda^\prime } 
+ \psi_{ii}^{ \lambda \lambda }  A_{ i \lambda^\prime }^2 
+ 4 \psi_{ii}^{ \lambda \lambda^\prime }  A_{ i \lambda } A_{ i \lambda^\prime } 
+ \psi_{ii}^{ \lambda^\prime \lambda^\prime } A_{ i \lambda }^2  
 + A_{ i \lambda }^2 A_{ i \lambda^\prime }^2 \right]  \\
&-
\frac{1}{2} \sum_{ij, \lambda \lambda^\prime} v_{ij}^{ \lambda \lambda^\prime } \left[ A_{i\lambda} A_{j \lambda^\prime} + \psi_{ij}^{\lambda \lambda^\prime} \right] -
 \Delta^2 \sum_{ i \lambda }  ( \DM_{ii}^{-1})^{ \lambda \lambda }- 
\sum_{i \lambda} E_{i \lambda }^0 A_{ i \lambda} \nonumber \\ 
&
- k_B T  N  \sum_{ {\bm q} \lambda } \left\{  \frac{\beta  \Omega_{ {\bm q} \lambda }}{2}     \coth\left( \frac{ \beta  \Omega_{ {\bm q} \lambda}}{2} \right) - \ln\left[ 2 \sinh\left( \frac{ \beta  \Omega_{ {\bm q} \lambda} }{ 2 } \right) \right] \right\}, \nonumber
\end{align} 
where, $\psi_{ij}^{\lambda \lambda^\prime} $ are temperature and disorder induced fluctuations of polarization
 between local soft mode components  $ Q_{i \lambda } $ and $  Q_{j \lambda^\prime } $, 
\begin{align}
\label{eq:psi}
\psi_{ij}^{\lambda \lambda^\prime} = \ovl{\bra \left( Q_{i \lambda} -  A_{i \lambda}  \right) \left( Q_{j \lambda^\prime} -  A_{j \lambda^\prime}  \right) \ket}
= \frac{1}{N} \sum_{ {\bm q},  \ovl{\lambda}  }   e^{- i {\bm q} \cdot {\bm R}_{ij}} \, 
(b_{\bm q}^{\dagger})_{  \lambda \ovl{\lambda} } \left( b_{\bm q} \right)_{ \ovl{\lambda} \lambda^\prime } 
 \left( \psi_{ \bm q}  \right) _{ \ovl{\lambda} },      
\end{align}
with Fourier component,
\begin{align*}
\left( \psi_{ \bm q}  \right)_{ \ovl{\lambda} } =  \frac{ 1 }{ 2  \Omega_{ {\bm q} \ovl{\lambda} } } \coth\left( \frac{ \beta  \Omega_{ {\bm q} \ovl{\lambda}}}{2} \right) + \frac{ \Delta^2 }{  \Omega_{ {\bm q} \ovl{\lambda} }^4 }.
\end{align*}
$(b_{\bm q})_{  \lambda \ovl{\lambda} } $ is a unitary  transformation that takes $ ( D_{\bm q} )^{ \lambda \lambda^\prime }  $ to its diagonal representation.

A standard procedure gives the following  dynamical matrix,
\begin{align}
\label{eq:DM}
\left( \DM_{\bm q} \right) _{\alpha \nu} 
& =  \left[ \kappa +  3( \gamma_1 - \gamma_2) \left( A_\alpha^2 + \psi_{0}^{\alpha \alpha}   \right)
    + \gamma_2  \sum_{\lambda} \left( A_\lambda^2 + \psi_{0}^{\lambda \lambda}   \right) \right] \delta_{\alpha \nu}  \nonumber  \\
   &~~~~~~~~~~~~~~~~~~~~~~~~~~~~~~~~~~~~~~~~~~~~~~~~~~~
   + 2 \gamma_2 \left( A_\alpha A_\nu + \psi_{0}^{\alpha \nu}  \right) - v_{\bm q}^{\alpha \nu},
\end{align}
where $  \psi_{0}^{\alpha \nu} $ is given in Eq. (\ref{eq:psi}). The diagonalization of $ \left( \DM_{\bm q} \right) _{\alpha \nu}  $ gives the squared of the soft phonon 
frequencies $\left( \Omega_{ {\bm q} \lambda } \right)^2$.

Minimization of  the free energy (\ref{eq:free1}) with respect to  $A_{i \lambda}$ gives the following result,
\begin{align}
\label{eq:p}
 \sum _{\nu=1}^3 \left[ \left( \DM_{{\bm q}=0} \right) _{\alpha \nu} - 2  \left( (\gamma_1 - \gamma_2) A_{\alpha}^2 + \gamma_2 \sum_{\lambda} A_\lambda^2  \right) \delta_{\alpha \nu}    \right]  A_{\nu} = E_{\alpha}^0 A_\alpha.
\end{align}
In writing Eq. (\ref{eq:p}) we have used the property that $v_{ij}^{\alpha \lambda} $ is translationally invariant
so  the summation $\sum_{i \lambda} v_{ij}^{\alpha \lambda} $ does not depend
on the origin $i$.  $\left( \DM_{{\bm q}=0} \right) _{\alpha \nu} $ depends on 
the direction in  which ${\bm q} \to 0$ because  $ v_{{\bm q}}^{\alpha \lambda} $ is non-analytic. Eqs. (\ref{eq:DM}) and (\ref{eq:p}) 
are the starting point of our analysis. 

We first consider the cubic phase. 
For the cubic phase,  there is no long-range order ($ A_x = A_y= A_z = 0$) 
and, by symmetry,  $ \psi_0 \equiv \psi_0^{xx} = \psi_0^{yy}=\psi_0^{zz},~ \psi_0^{xy}=\psi_0^{xz}=\psi_0^{yz}=0$~\cite{Pytte1972a}.  
Therefore, the dynamical matrix has the form,
\begin{align}
\label{eq:DMCubic}
\left( \mathcal{D}_{\bm q} \right) _{ \lambda \lambda^\prime}  = \left[ \kappa + \left( 3 \gamma_1 + 2 \gamma_2  \right) \psi_{0} \right]  \delta_{ \lambda \lambda^\prime }   + \left( v_{\bm q} \right) _{ \lambda \lambda^\prime}.
\end{align}
For an arbitrary direction of ${\bm q}$ the diagonalization of $ \left( \mathcal{D}_{\bm q} \right) _{ \lambda \lambda^\prime} $ gives
 a doubly degenerate transverse optic (TO) mode $ \Omega_{ {\bm q}}^\perp $, and a singlet longitudinal optic (LO) mode 
$  \Omega_{ {\bm q}}^\parallel $, given as follows,
\begin{subequations}
\label{eq:TOLOC}
\begin{align}
\left( \Omega_{ {\bm q}}^{\parallel} \right)^2  &= \left( \Omega_{\bm q}^{\perp} \right)^2  + C^2, \\
 \left( \Omega_{ {\bm q}}^{\perp} \right)^2 &=  \left( \Omega_{0}^{\perp} \right)^2 + B^2 \left| \bm{q} \right|^2, 
\end{align}
\end{subequations}
where, 
\begin{align}
\label{eq:TOLOC0}
 \left( \Omega_{0}^{\perp} \right)^2  &=  - \omega_0^2 + \left( 3 \gamma_1 + 2 \gamma_2  \right)  \psi_{0} ,
\end{align}
is the zone-center TO mode frequency and $\omega_0 \equiv \sqrt{v_0 - \kappa }$.
For cubic symmetry, the transformation matrix ${\bm b}_{\bm q}$ that diagonalizes the dynamical matrix (\ref{eq:DMCubic}) takes the form,
\begin{align*}
{\bm b} = 
\begin{pmatrix}
-\sin \phi  & \cos \theta  \cos \phi  & \cos \phi  \sin \theta  \\
 \cos \phi  & \cos \theta  \sin \phi  & \sin \theta  \sin \phi  \\
 0 & -\sin \theta  & \cos \theta  
\end{pmatrix},
\end{align*}
where $\theta$ and $\phi$ are the usual azimuthal and polar angles in spherical coordinates.

We now calculate the fluctuations $\psi_0$, 
\begin{align}  
\label{eq:Psiq}
 \begin{pmatrix}
   \psi_0^{xx} & \psi_0^{xy} & \psi_0^{xz} \\
   ... &  \psi_0^{yy} & \psi_0^{yz} \\
  ... &  ... &  \psi_0^{zz} 
    \end{pmatrix}     
 &= \frac{1}{N} \sum_{\bm q}   \begin{pmatrix}
   \psi_{\bm q}^{xx} & \psi_{\bm q}^{xy} & \psi_{\bm q}^{xz} \\
   ... &  \psi_{\bm q}^{yy} & \psi_{\bm q}^{yz} \\
  ... &  ... &  \psi_{\bm q}^{zz} 
    \end{pmatrix}    
= \frac{1}{N} \sum_{\bm q}
 {\bm b}_{\bm q}   
 \begin{pmatrix}
   \psi_{\bm q}^{\perp}  & 0 & 0 \\
   0 & \psi_ {\bm q}^{\perp}  & 0 \\
   0 &   0 &   \psi_{\bm q}^{\parallel} 
    \end{pmatrix}     
  {\bm b}_{\bm q}^T   \nonumber \\
  &= 
   \frac{1}{N} \sum_{\bm q} \begin{pmatrix}
  \psi_ {\bm q}^{\perp} c^2_\theta  c^2_\phi +   \psi_{\bm q}^{\parallel} s^2_\theta  c^2_\phi +   \psi_ {\bm q}^{\perp} s^2_\phi  & -( \psi_ {\bm q}^{\perp}-  \psi_{\bm q}^{\parallel}) c_\phi s^2_\theta s_\phi  & (  \psi_{\bm q}^{\parallel}-  \psi_ {\bm q}^{\perp} ) c_\theta  c_\phi  s_\theta \\
... &  \psi_ {\bm q}^{\perp} c^2_\phi +\left(    \psi_{\bm q}^{\parallel}c^2_\theta +  \psi_{\bm q}^{\parallel} s^2_\theta \right) s^2_\phi  & (  \psi_{\bm q}^{\parallel}- \psi_ {\bm q}^{\perp}) c_\theta s_\theta s_\phi \\
... & ... &   \psi_{\bm q}^{\parallel} c^2_\theta+ \psi_ {\bm q}^{\perp} s^2_\theta  \\
  \end{pmatrix}   
\end{align}
where $s_\phi \equiv \sin \phi, c_\phi \equiv \cos \phi, s_\theta \equiv \sin \theta, c_\theta \equiv \cos \theta$ and, 
\begin{subequations}
\label{eq:Psi}
\begin{align}
 \psi_{\bm q}^{\perp}  &= \frac{ 1 }{ 2  \Omega_{\bm q}^{\perp} } \coth\left( \frac{ \beta   \Omega_{ {\bm q}}^{\perp} }{2} \right) + \frac{\Delta^2}{ \left( \Omega_{ {\bm q}}^{\perp}  \right)^4 }, \\
 \psi_{\bm q}^{\parallel} &= \frac{ 1 }{ 2 \Omega_{ {\bm q}}^{\parallel} } \coth\left( \frac{ \beta  \Omega_{ {\bm q}}^{\parallel} }{2} \right) + \frac{\Delta^2}{ \left( \Omega_{ {\bm q}}^{\parallel}  \right)^4 }.
\end{align}
\end{subequations}
By taking the continuum limit over a sphere of wave-vector ${\bm Q}$ and calculating the angular integrals, we find the result,
\begin{align*}
 \begin{pmatrix}
   \psi_0^{xx} & \psi_0^{xy} & \psi_0^{xz} \\
   ... &  \psi_0^{yy} & \psi_0^{yz} \\
  ... &  ... &  \psi_0^{zz} 
    \end{pmatrix}     
    = 
    \frac{1}{Q^3 } \int_0^Q dq q^2 
    \begin{pmatrix}
       2 \psi_{\bm q}^{\perp} +\psi_{\bm q}^{\parallel}  & 0 & 0 \\
          0  &   2 \psi_{\bm q}^{\perp} + \psi_{\bm q}^{\parallel}  & 0  \\
           0  &   0 &  2 \psi_{\bm q}^{\perp} +\psi_{\bm q}^{\parallel} 
    \end{pmatrix}.
\end{align*}
Thus, 
\begin{align}
\label{eq:Psi0}
\psi_0 \equiv \psi_0^{xx} = \psi_0^{yy} = \psi_0^{zz} = \frac{1}{Q^3 } \int_0^Q dq q^2  \left(    2 \psi_{\bm q}^{\perp} +\psi_{\bm q}^{\parallel} \right).
\end{align}
Equations (\ref{eq:TOLOC}), (\ref{eq:TOLOC0}), (\ref{eq:Psi}), (\ref{eq:Psi0}) determine de temperature and disorder dependence
of the TO and LO mode frequencies for the cubic phase.

We now consider the rhombohedral phase. 
For the rhombohedral phase, we assume a homogenous order parameter along the cube diagonal,
$A_{x}^2 = A_{y}^2 = A_{z}^2 \equiv \frac{1}{3}A^2$. Also, by symmetry,  $ \psi_0^{11} \equiv \psi_0^{xx}=\psi_0^{yy}=\psi_0^{zz}, \psi_0^{12} \equiv \psi_0^{xy}=\psi_0^{xz}=\psi_0^{yz}$~\cite{Pytte1972a}.  
The dynamical matrix is as follows,
\begin{subequations}
\label{eq:DMR}
\begin{align}
\left( \mathcal{D}_{\bm q} \right)_{\alpha \alpha} &= \kappa + \left( 3 \gamma_1 + 2 \gamma_2 \right) \left( \frac{1}{3} A^2 + \psi_0^{11}   \right)- v_{\bm q}^{\alpha \alpha}, \\
 \left( \mathcal{D}_{\bm q} \right)_{\alpha \nu} &=   2 \gamma_2 \left(\frac{1}{3} A^2 + \psi^{12}_0    \right)   - v_{\bm q}^{\alpha \nu}, ~~\alpha \neq \nu. 
\end{align}
\end{subequations}
where $~~\alpha, \nu=x,y,z$. We first  identify the soft mode frequencies. Pure longitudinal and transverse modes are obtained for wavevectors in the $(111)$ direction and 
the plane transverse to it. For $ {\bm q} \perp (1,1,1) $ diagonalization of the 
dynamical matrix gives  two distinct TO mode frequencies 
$\Omega_{0 1}^{\perp} $ and  $\Omega_{0 3}^{\perp} $ and one LO frequency $ \Omega_{0 1}^{\perp} + 3 (C^2/3)$ at the zone-center.
For $ {\bm q} \parallel (1,1,1) $ there is  a  doubly degenerate TO mode frequency $\Omega_{0 1}^{\perp}  $ and one LO mode  frequency $ \Omega_{0 3}^{\perp}+ 3 (C^2/3)$ at the zone-center. $ \Omega_{0 1}^{\perp}  $ and $ \Omega_{0 3}^{\perp}  $ are given as follows,
\begin{subequations}
\label{eq:omegarhomboq1}
\begin{align}
\left( \Omega_{01}^\perp \right)^2  &=  - \omega_0^2 + \left( 3 \gamma_1 + 2 \gamma_2 \right) \left( \frac{1}{3} A^2 + \psi_0^{11}  \right) - 2 \gamma_2 \left( \frac{1}{3} A^2+ \psi^{12}_0  \right) , \\
 \left( \Omega_{03}^\perp \right)^2 &=   - \omega_0^2 + \left( 3 \gamma_1 + 2 \gamma_2 \right) \left( \frac{1}{3} A^2 + \psi_0^{11}  \right) + 4 \gamma_2 \left( \frac{1}{3} A^2+ \psi^{12}_0  \right).
\end{align}
\end{subequations}
While exact expressions can be derived for the phonon dispersions 
from the dynamical matrix $(\ref{eq:DMR})$, they are  too elaborated and not enlightening. 
Instead we calculate them from perturbation theory. 
Our unperturbed basis is that of the cubic phase, therefore making the frequency splitting $\Omega_{0 3}^{\perp}-  \Omega_{0 1}^{\perp} $ 
the expansion parameter. It is also convenient to write the wavevector as  ${\bm q} = q_L \hat{\bm q}_{L} +  q_{T1} \hat{\bm q}_{T1} + q_{T2} \hat{\bm q}_{T2}$, 
where $ \left\{ \hat{\bm q}_{L}, \hat{\bm q}_{T1}, \hat{\bm q}_{T2} \right\} $  is a right-handed coordinate system 
where  $ \hat{\bm q}_{L} $ is along the $(111)$ direction and $\hat{\bm q}_{T1,2}$ are transverse to it. 
The result is as follows,
\begin{subequations}
\label{eq:omegarhomboq2}
\begin{align}
 \left( \Omega_{ {\bm q} 1 } \right)^2 &= \left( \Omega_{0 1}^{\perp} \right)^2  + B^2 \left| {\bm q}  \right|^2,  \\
 \left( \Omega_{ {\bm q} 2 } \right)^2 &= \left( \Omega_{0 1}^{\perp} \right)^2  + B^2 \left| {\bm q}  \right|^2   +  \left[  \left( \Omega_{0 3}^{\perp} \right)^2  - \left( \Omega_{0 1}^{\perp} \right)^2   \right]    \frac{ q_{T}^2 }{ \left| {\bm q}  \right|^2 }, \\
 \left( \Omega_{ {\bm q} 3 } \right)^2 &= \left( \Omega_{0 1}^{\perp} \right)^2  + B^2 \left| {\bm q}  \right|^2 + 3 (C^2/3)   + \left[ \left( \Omega_{0 3}^{\perp} \right)^2  - \left( \Omega_{0 1}^{\perp} \right)^2   \right]   \frac{  q_L^2 }{ \left| {\bm q}  \right|^2 },
\end{align}
\end{subequations}
where $ q_T^2 = q_{T1}^2+q_{T2}^2 $ and with a transformation matrix given by,
\begin{align*}
{\bm b} = 
\begin{pmatrix}
\frac{1}{\sqrt{6}} & \frac{1}{\sqrt{2}} & \frac{1}{\sqrt{3}} \\
\frac{1}{\sqrt{6}} & -\frac{1}{\sqrt{2}} &  \frac{1}{\sqrt{3}} \\
-\frac{2}{\sqrt{6}} & 0 &   \frac{1}{\sqrt{3}} \\
\end{pmatrix}
\begin{pmatrix}
-\sin \phi  & \cos \theta  \cos \phi  & \cos \phi  \sin \theta  \\
 \cos \phi  & \cos \theta  \sin \phi  & \sin \theta  \sin \phi  \\
 0 & -\sin \theta  & \cos \theta  
\end{pmatrix},
\end{align*}
For an applied field $(E_0/\sqrt{3})(1,1,1)$, minimization of the free energy with respecto to the order parameter gives the following result,
\begin{align*}
\left[ \left( \mathcal{D}_{\bm 0} \right) _{11} + 2\left(  \mathcal{D}_{\bm 0} \right) _{12} - \frac{2}{3}\left( \gamma_1 + 2\gamma_2 \right)A^2  \right] A =E_0,
\end{align*}
which can be rewritten in terms of the soft mode frequencies as follows,
\begin{align}
\label{eq:OPR}
\left[ { \Omega_{03}^\perp }^2   - \frac{2}{3}\left( \gamma_1 + 2 \gamma_2 \right)A^2  \right] A =E_0.
\end{align}
We now calculate $  \psi_0^{xx} $ and $ \psi_0^{xy}$,
\begin{align*}
 \begin{pmatrix}
   \psi_0^{xx} & \psi_0^{xy} & \psi_0^{xz} \\
    ... &  \psi_0^{yy} & \psi_0^{yz} \\
    ... & ... &  \psi_0^{zz} 
    \end{pmatrix}     
&= \frac{1}{N} \sum_{\bm q}
 {\bm b}_{\bm q}   
 \begin{pmatrix}
   \psi_{ {\bm q} 1} & 0 & 0 \\
   0 &  \psi_{ {\bm q} 2} & 0 \\
   0 &   0 &   \psi_{ {\bm q} 3} 
    \end{pmatrix}   
  {\bm b}_{\bm q}^T,     
\end{align*}
where
\begin{subequations}
\label{eq:PsiRq}
\begin{align}
 \psi_{ {\bm q} 1} &= \frac{ 1 }{ 2  \Omega_{ {\bm q} 1 } } \coth\left( \frac{ \beta  \Omega_{ {\bm q}1 }}{2} \right) + { \frac{\Delta^2}{ \left( \Omega_{ {\bm q}1 } \right)^4}},  \\
 \psi_{ {\bm q} 2} &= \frac{ 1 }{ 2 \Omega_{ {\bm q} 2 } } \coth\left( \frac{ \beta  \Omega_{ {\bm q}2 }}{2} \right)+  \frac{\Delta^2}{ \left( \Omega_{ {\bm q}2 } \right)^4},  \\
 \psi_{ {\bm q} 3} &= \frac{ 1 }{ 2 \Omega_{ {\bm q} 3 } } \coth\left( \frac{ \beta  \Omega_{ {\bm q} 3 }}{2} \right)+   \frac{\Delta^2}{ \left( \Omega_{ {\bm q}3 } \right)^4}.  
\end{align}
\end{subequations}

To proceed further, we the continuum limit as we did  in the previous section and calculate the
integrals over  $\phi$. The result is the following,
\begin{subequations}
\label{eq:PsiR0}
\begin{align}
   \psi_0^{11} = \psi_0^{xx} =  \psi_0^{yy} =  \psi_0^{zz} &=  \frac{1}{Q^3}  \int_0^Q dq q^2 \int_0^\pi d \theta \sin \theta \frac{1}{2} ( \psi_{ {\bm q} 1} + \psi_{ {\bm q} 2} +    \psi_{ {\bm q} 3} ),    \\
    \psi_0^{12} = \psi_0^{xy} = \psi_0^{xz} = \psi_0^{yz} &=  \frac{1}{Q^3}  \int_0^Q dq q^2 \int_0^\pi d \theta \sin \theta  \frac{1 }{8} ( -2 \psi_{ {\bm q} 1} +\psi_{ {\bm q} 2} + \psi_{ {\bm q} 3} - 3 \left[\psi_{ {\bm q} 2} - \psi_{ {\bm q} 3} \right] \cos 2 \theta ).  
 \end{align} 
  \end{subequations}
Equations (\ref{eq:omegarhomboq1})-(\ref{eq:PsiR0}) determine de temperature and disorder dependence
of the order parameter $A$ and the TO and LO mode frequencies in the rhombohedral phase.

\section{Acknowledgements}
GGGV acknowledges useful discussions with Peter B. Littlewood.

\section{Competing Interests} 
The authors declare no competing financial interests.

\section{Author Contributions} 
GGGV conceived the study and wrote the manuscript; 
JRAG and GGGV performed the calculations and discussed the results and implications at all stages.

\section{Funding}
Work at the University of Costa Rica is supported by the Vice-rectory for Research under 
the project no. 816-B7-601.
Work at Argonne is supported by U.S. Department of Energy, Office of Basic Energy Sciences
under contract no. DE-AC02-06CH11357.

\end{document}